# Molecularly-thin Anatase field-effect transistors fabricated through the solid state transformation of titania nanosheets†

S. Sekizaki,[a] M. Osada[b] and K. Nagashio[*a,c]

We demonstrate the field-effect transistor (FET) operation of molecularly-thin anatase phase produced through solid state transformation from $Ti_{0.87}O_2$ nanosheets. Monolayer $Ti_{0.87}O_2$ nanosheet with a thickness of 0.7 nm is two-dimensional oxide insulators in which Ti vacancies are incorporated, rather than oxygen vacancies. Since the fabrication method, in general, largely affects the film quality, the anatase films derived from $Ti_{0.87}O_2$ nanosheet show interesting characteristics, such as no photocurrent peak at ~2 eV, which is related to oxygen vacancies, and a larger band gap of 3.8 eV. The 10-nm thick anatase FETs exhibit superior transport characteristics with a maximum mobility of ~1.3 $cm^2V^{-1}s^{-1}$ and current on/off ratio of ~$10^5$ at room temperature. The molecularly-thin anatase FET may provide new functionalities, such as field-effect control of catalytic properties.

## Introduction

Although oxide field effect transistors (FETs), such as InGaZnO, ZnO, $InO_x$, $SnO_2$, and $TiO_x$, have attracted great interest mainly in flexible and transparent display applications,[1,2] the addition of other functionalities will widen their practical application, especially for the recent great development of sensors for the Internet of Things (IoT). $TiO_x$ is one of the most promising candidates because of its photocatalytic properties. Ultraviolet (UV) irradiation excites $e$-$h$ pairs, causing a redox reaction on the $TiO_x$ surface.[3] Recently, the superconductivity[4], as well as a change in magnetic anisotropy[5], has been shown to be induced by the electric field effect. If it is possible to replace UV irradiation by the electric field effect,[6] the range of applications will be expanded. Although $TiO_x$ field effect transistors (FETs) have been reported with a backgate structure in which the $TiO_x$ surface is available,[6-11] it is difficult to induce carriers on the $TiO_x$ surface, rather than at the interface between $TiO_x$ and the gate insulator, because the thickness of the $TiO_x$ layer is generally thicker than 20 nm in order to retain crystal quality.

Here, we utilized $Ti_{0.87}O_2$ nanosheets as a template for solid state fabrication of molecularly-thin anatase films. $Ti_{0.87}O_2$ nanosheet is a new class of titania nanomaterials, derived from a layered titanate ($K_{0.8}Ti_{1.73}Li_{0.27}O_4$) via soft chemical reaction.[12-14] Alkali metal ions K and Li are extracted in acid treatment during delamination process, and the original layered structure is retained in $Ti_{0.87}O_2$ nanosheets. Then, $Ti_{0.87}O_2$ nanosheets is delaminated with a tetrabutylammonium (TBA) aqueous solution. $Ti_{0.87}O_2$ nanosheets are insulators with a band gap of 3.8 eV[15] and a high dielectric constant of ~125[16], as the original layered structure is retained by introducing 13% Ti vacancies instead of oxygen vacancies.[17,18] Moreover, the nanosheet transforms into the highly-crystalline, $c$-axis oriented anatase phase with increasing annealing temperature.[19]

Therefore, it may be possible to fabricate anatase $TiO_2$ FETs with thicknesses of less than 10 nm through solid state transformation from $Ti_{0.87}O_2$ nanosheets. Since it is well recognized that the $TiO_x$ deposition conditions largely affect the film quality, a starting material without oxygen vacancies may result in improved properties. In this study, anatase thin-film FETs are fabricated from $Ti_{0.87}O_2$ nanosheets, and their transport properties are discussed.

## Experimental

A colloidal suspension of $Ti_{0.87}O_2$ nanosheet was spin-coated on a $SiO_2$ (90 nm)/$n^+$-Si substrate at 3000 rpm. TBA was removed by ultraviolet light (UV) irradiation for 24 hours,[20] where the sample was unintentionally heated to ~150 °C by the UV irradiation. Monolayer $Ti_{0.87}O_2$ nanosheets with a lateral size of ~5 μm and a thickness of ~1 nm were found by optical microscopy, as the thickness of $SiO_2$ was adjusted to ~90 nm in order to maximize the optical contrast by thin-film interference, referring to the graphene observation.[21] Annealing to induce phase transformation from $Ti_{0.87}O_2$ nanosheets to the anatase phase was carried out in a tube furnace under $O_2$ gas flow for 1 hour at different temperatures. Then, the phase was characterized by micro-Raman spectroscopy with an Ar laser of 488 nm and ~0.3 mW and photoluminescence spectroscopy (PL) with a HeCd laser of 325 nm and ~0.5 mW. For FET device fabrication, the electrode pattern was drawn by the conventional electron beam lithography technique.[21] Al/Au multi-terminal electrodes were deposited by the thermal evaporation. The contact metal was Al, while Au was used to prevent the oxidation of Al. The electric transport measurement was carried out over a temperature range of 100 - 350 K in vacuum.

[a.] Department of Materials Engineering, The University of Tokyo, Tokyo 113-8656, Japan. Email: nagashio@material.t.u-tokyo.ac.jp
[b.] National Institute of Materials Science, Ibaraki 305-0044, Japan
[c.] PRESTO, Japan Science and Technology Agency (JST), Tokyo 113-8656, Japan

†Electronic Supplementary Information (ESI) available: [EBSD analysis].

## Results and discussion

**Phase transformation from $Ti_{0.87}O_2$ nanosheets to anatase phase**

Fig. 1(a) shows an optical microscopy image with Nomarski interference contrast for a spin-coated sample after annealing. The $Ti_{0.87}O_2$ nanosheets are classified into three types based on the sample morphology: monolayer, multilayer (monolayer $Ti_{0.87}O_2$ nanosheets were randomly laminated without crystal orientation matching during spin-coating), and bulk (it was not exfoliated in the $Ti_{0.87}O_2$/TBA suspension. Therefore, $Ti_{0.87}O_2$ nanosheets are stacked with retaining the crystal orientation). The extinction coefficient in the visible range is ideally zero for $Ti_{0.87}O_2$ nanosheets due to their wide band gap.[15,22] Therefore, the Nomarski technique is required to enhance the contrast. Although the $Ti_{0.87}O_2$ nanosheets were unintentionally heated to ~150 °C during UV irradiation, they did not transform to anatase, as confirmed by Raman spectrum.

The samples were annealed at temperatures ranging from 400 °C to 900 °C for 1 hour under $O_2$ gas flow to facilitate solid state phase transformation from $Ti_{0.87}O_2$ nanosheets to the anatase phase. Although the $Ti_{0.87}O_2$ nanosheets do not show any Raman peaks in the range of 100 - 450 $cm^{-1}$ due to the small Raman scattering efficiency, the main peak of the anatase phase, $E_g(1)$, appears at ~142 $cm^{-1}$ after annealing at high temperatures.[23,24] Fig. 1(b) shows the intensity ratio of $E_g(1)$ and the Si peak at 300 $cm^{-1}$ ($Si_{300}$) and the full width at half maximum (FWHM) of $E_g(1)$ for the multilayer sample. The intensity ratio and FWHM show a maximum and minimum at 800 °C, respectively. These temperature dependence was consistent with the previous X-ray diffraction analysis for the annealing experiment of wafer-scale multilayer $Ti_{0.87}O_2$ nanosheets stacked by Langmuir-Blodgett method.[19] Based on this experiment, the annealing temperature of 800 °C was selected for subsequent characterization and device fabrication. This choice is also supported by a recent report in which both Ti-Si interdiffusion and intermediate phase formation at the $TiO_x/SiO_2$ interface is absent even at the annealing temperature of 800 °C.[11]

The Raman spectra of the three types of samples annealed at 800 °C are compared in Fig. 1(c), where the Raman spectrum of the $SiO_2$/Si substrate is also shown for reference. The main peak of the anatase phase, $E_g(1)$, is observed at 142 $cm^{-1}$ for the multilayer and bulk samples, but not for the monolayer sample. The anatase peak intensity of $B_{1g}(1)$, as well as $E_g(1)$, for the bulk sample is higher than that for the multilayer sample, suggesting that the crystallinity of the anatase phase in the bulk sample is better than that in the multilayer

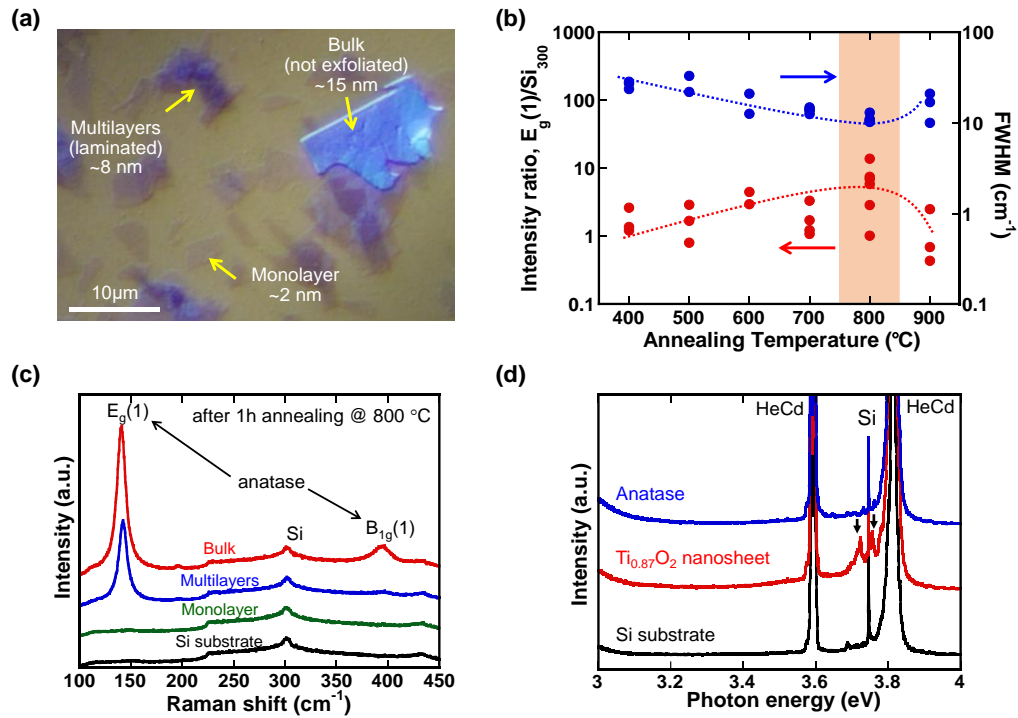

**Fig. 1** (a) Optical microscopy image with Nomarski interference contrast for $Ti_{0.87}O_2$ nanosheets. (b) Raman intensity ratio for $E_g(1)/Si_{300}$ and FWHM of $E_g(1)$ as a function of annealing temperature. (c) Raman spectra for the three types of anatase samples, along with the spectrum of the Si substrate as a reference. (d) PL spectra for the bulk samples of anatase $TiO_2$ and $Ti_{0.87}O_2$ nanosheets, along with the spectrum of the Si substrate as a reference. Two small PL peaks, indicated by arrows, come from $Ti_{0.87}O_2$ nanosheets. It should be noted that the peaks at ~3.82 eV, 3.74 eV, and 3.6 eV with high intensities result from the Rayleigh peak of the HeCd laser, the Si Raman spectrum, and the bright line of the HeCd laser.

sample. The crystallinity was analysed in more detail by electron backscattering diffraction (EBSD), as shown in supplemental **Fig. S1**. The bulk sample shows a clear single crystalline nature, while multilayer and monolayer samples are polycrystalline and amorphous, respectively. At least two layers of nanosheets are required to satisfy the periodicity of the anatase crystal structure, and thus a monolayer is difficult to transform into the anatase phase. These features are consistent with the previous annealing experiment.[19]

In the PL spectrum in **Fig. 1(d)**, measured at room temperature, small two peaks near 3.8 eV observed in the parent compound ($H_{1.07}Ti_{1.73}O_4 \cdot H_2O$) of $Ti_{0.87}O_2$ nanosheets[25], indicated by arrows, disappears after annealing, because the anatase phase is a semiconductor with an indirect gap of 3.2 eV.[26] It should be noted that the peaks at ~3.82 eV, 3.74 eV, and 3.6 eV with high intensities result from the Rayleigh peak of the HeCd laser, the Si Raman spectrum, and the bright line of the HeCd laser. This also confirms the solid state phase transformation from $Ti_{0.87}O_2$ nanosheets to the anatase phase.

**FET characteristics**

**Fig. 2(a)** shows the transfer characteristics of anatase FETs measured at room temperature in a vacuum

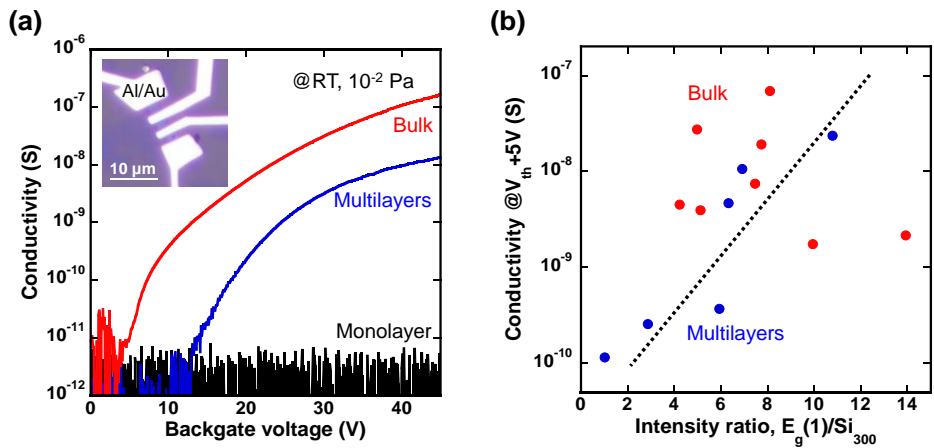

**Fig. 2** (a) FET characteristics of the multilayer and bulk samples. Inset: optical image of a typical bulk anatase $TiO_x$ FET, obtained by four-probe measurement. (b) Conductivity as a function of $B_{1g}(1)/Si_{300}$.

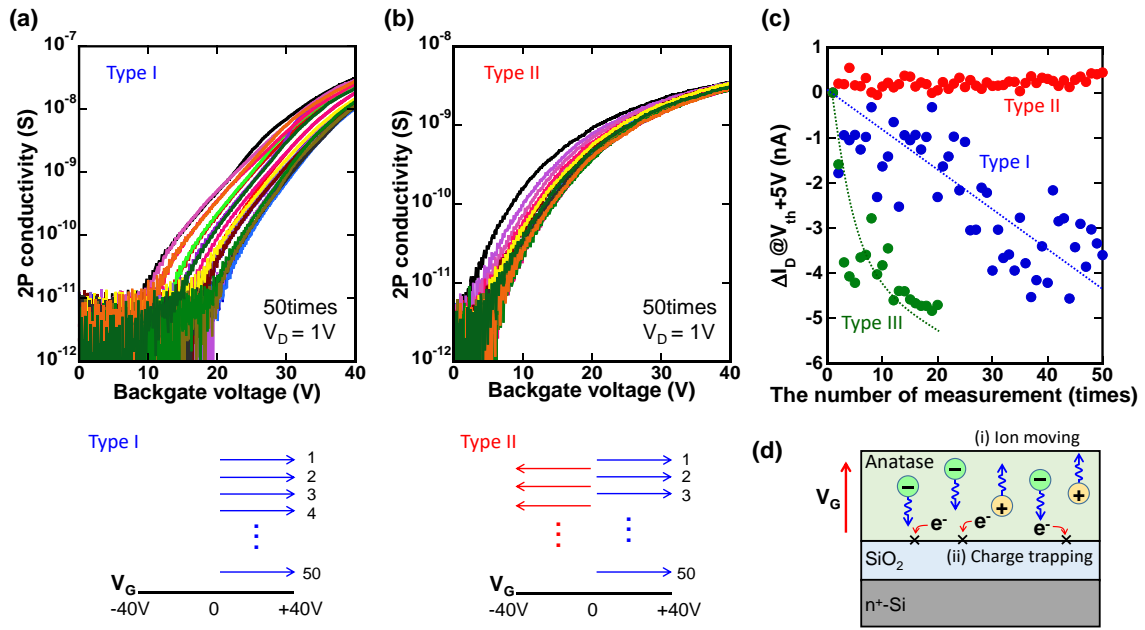

**Fig. 3** (a) Degradation behavior when only positive $V_G$ is applied over 50 cycles (type I measurement). (b) Degradation behavior when positive and negative $V_G$ is iteratively applied over 50 cycles (type II measurement). (c) Drain current reduction from the initial level as a function of the number of $IV$ measurements for the three kinds of measurements. (d) Schematic of the two origins of degradation, ion migration and charge trapping at the interface.

of ~$10^{-2}$ Pa. The $n^+$-Si substrate was used as the global backgate. No current modulation was observed in the monolayer sample, while the multilayer and bulk samples with thicknesses of ~8 - 15 nm exhibited clear transistor operation. As shown in **Fig. 2(b)**, there is a clear relationship between the conductivity at $V_G = V_{th} + 5$ V and the $E_g(1)/Si_{300}$ intensity ratio, where $V_G$, and $V_{th}$ are the backgate voltage and threshold voltage, respectively. The FET performance strongly depends on the crystallinity. The best values for the field effect mobility and the current on/off ratio of the bulk sample, analyzed by four-probe measurement, are ~1.3 $cm^2V^{-1}s^{-1}$ and ~$10^5$, respectively. Although these values are comparable with those in previous literature,[7-10] current modulation of the anatase $TiO_x$ FET was achieved even for a channel thickness of less than 10 nm in this study. This will facilitate carrier modulation at the anatase surface by the electric field effect.

**Degradation and recovery of FET characteristics**

Although FET modulation was achieved, degradation of the FET properties was observed when the measurement was repeated at positive $V_G$ (type I), as shown in **Fig. 3(a)**. $V_{th}$ gradually shifts to positive $V_G$. Moreover, the reduction in conductivity ($\Delta I_D$) from the initial current level, which was calculated at $V_G = V_{th} + 5$ V, is plotted as a function of the number of measurements in **Fig. 3(c)**. Similar degradation has been reported for a RF-sputtered $TiO_x$ anatase FET[27] and is often observed for other oxide transistors.[28] To separate the contribution from drain voltage ($V_D$) and $V_G$, $V_G$ was applied from -20 V to 40 V five times to the FET device without applying $V_D$, and then the transfer curve was measured at $V_D = 1$ V. This is defined as a single measurement and was repeated 20 times (type III measurement). The reduction in $\Delta I_D$ during type III measurements is more serious than that during type I measurements, suggesting that $V_G$ is more responsible for degradation. Since the application of positive $V_G$ results in a reduction in $\Delta I_D$, positive and negative $V_G$ were iteratively applied (type II measurement), as shown in **Fig. 3(b)**. In this case, the reduction in $\Delta I_D$ was clearly suppressed. Based on these experiments, the degradation is suggested to be caused by charge trapping at the $TiO_2/SiO_2$ interface[27] and/or the migration of ions in the $TiO_x$ channel[29] (**Fig. 3(d)**).

To understand the degradation behavior in more detail, the drain current ($I_D$) was measured over time at constant $V_G = 40$ V and $V_D = 1$ V, as shown in **Fig. 4(a)**. $I_D$ drastically decreased within several seconds and then gradually decreased. By curve fitting using the equation $I_D = a_1 e^{-t/\tau_1} + a_2 e^{-t/\tau_2} + a_3 e^{-t/\tau_3}$, three time constants were extracted, $\tau_1 =$ ~2.6 s, $\tau_2 =$ ~100 s, and $\tau_3 =$ ~1100 s. These time constants seem to be large for electron or hole trapping. Therefore, degradation may be dominated by ion migration. One possibility is the migration of oxygen vacancies, because the electromigration of oxygen vacancies has been suggested in $TiO_x$ previously.[29,30]

If the origin of degradation is the migration of oxygen vacancies, defect levels related to oxygen vacancies will be detected in the photocurrent spectrum.[11] Therefore, single energy photons were used to irradiate the anatase FET at $V_D = 1$ V and $V_G = 0$ V, and the photoexcited electrons were collected. **Fig. 4(b)** shows the photocurrent spectra as a function of photon energy for the multilayer and bulk anatase FETs. The broad photocurrents are detected between 1.5 to 2.5 eV in the multilayer sample, which is similar to the oxygen-vacancy-related photocurrent peak at around ~2 eV observed for the typical anatase FET fabricated by pulsed laser deposition.[11] The anatase phase in the multilayer sample is formed from the random stacking of monolayer $Ti_{0.87}O_2$ nanosheets, as schematically shown in **supplemental Fig. S1**. Therefore, long-range diffusion is

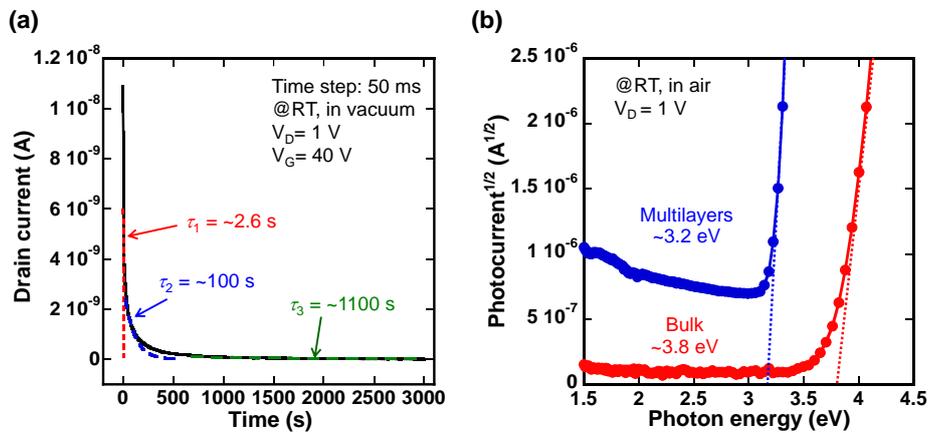

**Fig. 4** (a) Drain current as a function of time for the bulk sample. Dotted lines are the fitted curves. The time constants are shown in the figure. (b) Photocurrent spectra for the multilayer and bulk samples. The photocurrent measurement was carried out at $V_D = 1$ V and $V_G = 0$ V, that is, the FET was in the off-state.

required to transform the $Ti_{0.87}O_2$ nanosheets into an anatase crystal,[19] and oxygen vacancies may be introduced in the polycrystalline anatase crystals during transformation. On the other hand, it is quite interesting that there are no obvious photocurrent peaks for the bulk sample. It is reported that the atomic arrangement in the typical anatase is quite similar to that for $Ti_{0.87}O_2$ nanosheets stacked with retaining the crystal orientation.[19] Therefore, the bulk anatase phase is single crystalline in nature, as observed by EBSD. It is expected that the oxygen vacancy is be introduced during the transformation due to the quite short range diffusion. Although the species that undergo ion migration has not been specified at present, it is suggested that the migration of oxygen vacancies is negligible in bulk anatase, supported by no detection of photocurrent within the gap.

Moreover, the band gap energy ($E_G$) can be extracted from the photocurrent spectra. $E_G$ for multilayer anatase is ~3.2 eV, which is consistent with the literature data obtained from optical absorption spectra for typical anatase.[22] This also suggests that multilayer anatase includes oxygen vacancies. On the other hand, $E_G$ = ~3.8 eV for bulk anatase is much larger than that for typical anatase but is similar to $E_G$ = ~3.8 eV for $Ti_{0.87}O_2$ nanosheets, obtained by UV-vis absorption spectra.[15] The large $E_G$, as well as negligible oxygen vacancies, of the bulk sample suggests that the physical properties of the anatase phase can be controlled by the selection of starting material before solid state transformation. Therefore, the FET operation of the bulk $TiO_x$ is better than that of multilayer $TiO_x$.

**Intrinsic transport properties and estimation of Schottky barrier height**

The intrinsic transport properties of the bulk anatase FET can be discussed by applying type II measurements (the iterative $V_G$ method). **Fig. 5(a)** shows the temperature dependence of the conductivity obtained from the four-probe device. The typical field effect mobilities calculated at the maximum of transconductance for the three different single crystalline bulk anatase phases are shown as a function of temperature in **Fig. 5(b)**. The mobility data vary quite widely in the range of ~$10^{-3}$ to ~$10^0$ $cm^2V^{-1}s^{-1}$ due to the variation in crystallinity. When the mobility at room temperature is lower than 1 $cm^2$/Vs, the mobility decreases with decreasing temperature. From the analysis

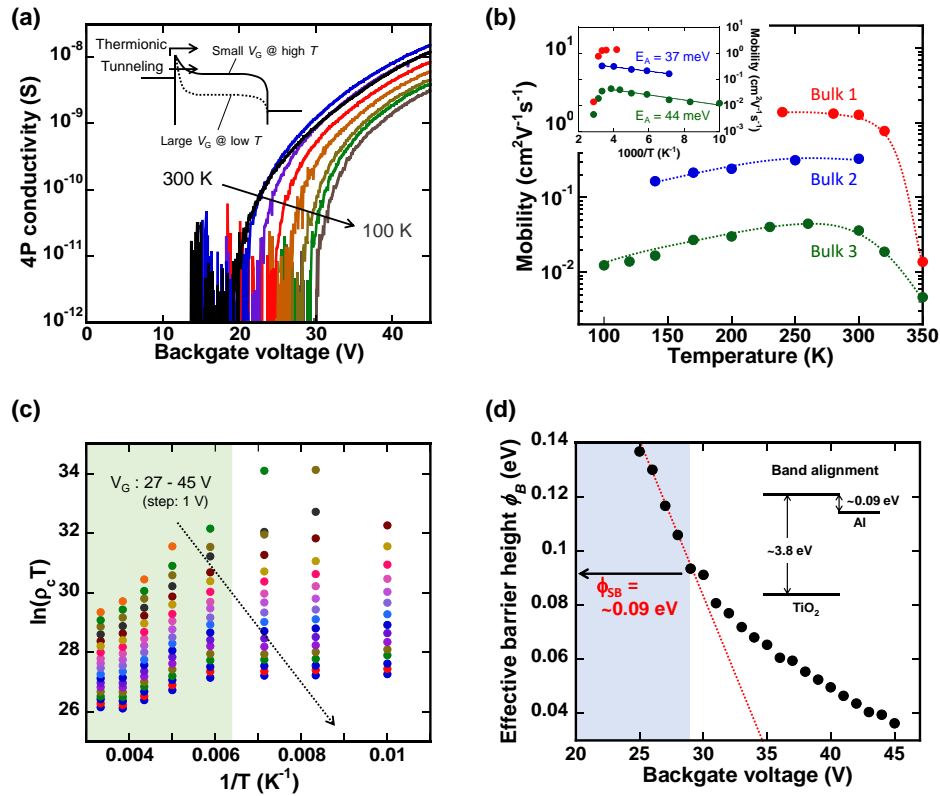

**Fig. 5** (a) Temperature dependence of the FET characteristics. The measured temperatures are 100, 120, 140, 170, 200, 230, 260, and 300 K. The inset shows the schematic of the Schottky barrier-type transistor. (b) Mobilities for three different devices based on the bulk samples. (c) Arrhenius plot of contact resistivity at different gate voltages. (d) Extracted effective barrier height $\phi_B$ as a function of backgate voltage. (d) Estimated band alignment for bulk anatase $TiO_2$ and Al.

in the inset, the activation energies are extracted as 37 and 44 meV for two cases, which suggests a hopping-type conduction mechanism at the anatase/$SiO_2$ interface and/or in the anatase crystal.[31] On the other hand, for the sample with a mobility higher than 1 $cm^2V^{-1}s^{-1}$, the mobility seems to increase with decreasing temperature. The theoretical maximum mobility at room temperature is calculated to be ~60 $cm^2V^{-1}s^{-1}$,[32] which is limited by phonon scattering. The present mobility data seem to move into phonon dominant transport. When the crystallinity is improved further, the transition from hopping-type conduction to conduction over fully delocalized energy bands will be clearly seen.

Another characteristic in **Fig. 5(a)** is that $V_{th}$ shifts to positive $V_G$ with decreasing temperature. In anatase FETs, metal contacts are used as the source and drain. Thus, anatase $TiO_x$ FETs are expected to be Schottky barrier-type transistors, as is generally the case for many organic,[33] oxide,[34] carbon nanotube[35] and two-dimensional layered semiconductors,[36] unlike the inversion-type transistors for conventional Si-MOSFETs with *p-n* junctions. With decreasing temperature, the thermionic emission current over the Schottky barrier decreases, and the tunneling current becomes dominant. Therefore, at lower temperatures, $V_G$ must be further applied to increase the tunneling current by decreasing the Schottky barrier width. This is the case for the present $V_{th}$ shift as a function of temperature, as schematically shown in the inset of **Fig. 5(a)**.

Here, the Schottky barrier height ($\phi_{B0}$) between the bulk anatase phase and the Al contact was estimated. Current transport processes through the metal/semiconductor junction are classified into three types: thermionic emission (TE), thermionic field emission (TFE), and field emission (FE).[37] A simple criterion to determine the dominant process can be set by $E_{00}/k_BT$, where $E_{00}$ is defined as $E_{00} = \frac{q\hbar}{2}\sqrt{\frac{n_{3D}}{m^*\varepsilon_s}}$, where $\hbar$, $m^*$, and $\varepsilon_s$ are the reduced Planck constant, the effective mass, and the electric permittivity of the semiconductor, respectively. The three-dimensional carrier density ($n_{3D}$) can be roughly estimated as $n_{3D} = n_{2D}$/thickness.[37] TE dominates for $E_{00}/k_BT \ll 1$, while TFE becomes the main process for $E_{00}/k_BT \gg 1$. For the anatase FET, this simple criterion suggests that TE is the main transport process because $E_{00}/k_BT \approx 10^{-2}$ at all temperatures over the range of 100 - 300 K. In case of TE, the specific contact resistivity is given by

$$\rho_c = \frac{k_B}{A^{**}Tq}\exp\left(\frac{q\phi_B}{k_BT}\right), \quad [1]$$

where $A^{**}$, $q$, and $\phi_B$ are the modified Richardson constant, the elementary charge, and the effective barrier height, respectively.[37-39]

**Fig. 5(b)** shows the Arrhenius plot of the contact resistivity at different $V_G$, estimated through the four-probe measurement. The effective barrier height can be calculated from the slope ($q\phi_B/k_B$). Two different slopes are clearly seen at the boundary temperature of 170 K (1/T = 0.00588), suggesting that TE is only applicable to the high-temperature region. Although the contact resistivity estimated from the present four-probe measurement (in units of $\Omega\mu$m) is different from the specific contact resistivity (in units of $\Omega\mu m^2$), Eq. [1] is used to estimate $\phi_B$ because only the slope is considered. Then, $\phi_B$ is plotted as a function of $V_G$ in **Fig. 5(c)**. The effective barrier height means the barrier height the carriers feel at a certain $V_G$, and the Schottky barrier height can be defined as the effective barrier height at the flat band voltage ($V_{FB}$). Since the linear relation of $\phi_B = \phi_{B0} - \gamma(V_G - V_{FB})$ exists when $V_G < V_{FB}$,[35] the Schottky barrier height is estimated to be ~0.09 eV for Al at $V_{FB}$ = 32 V. Finally, the band alignment for the Schottky junction of the bulk anatase phase and Al was experimentally determined, as shown in the inset of **Fig. 5(d)**. For Al, the present $\phi_{B0}$ value is smaller than $\phi_{B0}$ = ~0.3 eV, as given in previous reports.[31,40] Although the origin of the ohmic contact between $TiO_x$/Al has been discussed to be the increase in oxygen vacancies at the $TiO_x$/Al interface due to chemical reactions expected from the small electronegativity of Al,[40] the interfacial conditions in the present bulk anatase phase may be different due to its different physical properties. Further investigation is required.

## Conclusions

In this work, single crystalline anatase FETs were fabricated through solid state transformation from $Ti_{0.87}O_2$ nanosheets. Since the present anatase phase was fabricated from $Ti_{0.87}O_2$ nanosheets that did not have oxygen vacancies, there is no photocurrent peak at ~2 eV, which is related to oxygen vacancies, and the band gap of 3.8 eV is larger than that of typical anatase crystals. As for the transport characteristics, a maximum mobility of 1.3 $cm^2V^{-1}s^{-1}$ and current on/off ratio of ~$10^5$ were estimated at room temperature, and the temperature dependence of the mobility suggests the phonon dominant transport. Al contact to the present anatase phase is suggested to be reasonable, based on the Schottky barrier height analysis ($\phi_{B0}$ =0.09 eV). The ~10 nm thickness of the present anatase channel may provide other potential functionalities, such as control of redox reactions on the anatase surface by the field effect.

**Acknowledgements**

The authors would like to thank Dr. T. Yajima for fruitful discussion. This research was partly supported by the JSPS Core-to-Core Program, A. Advanced Research

Networks, JSPS KAKENHI Grant Numbers JP25107004, JP16H04343, JP16K14446, & JP26886003, JST PRESTO Grant Number JPMJPR1425, Japan.

**References**

1. E. Fortunato, P. Barquinha and R. Martins, *Adv. Mater.*, 2012, **24**, 2945-.
2. J. -S. Park, H. Kim and I. -D. Kim, *J. Electroceram.*, 2014, **32**, 117-.
3. A. Fujishima, X. Zhang and D. A. Tryk, *Surf. Sci. Rep.*, 2008, **63**, 515-582.
4. K. Ueno, S. Nakamura, H. Shimotani, H. T. Yuan, N. Kimura, T. Nojima, H. Aoki, Y. Iwasa and M. Kawasaki, *Nature nanotechnol.*, 2011, **6**, 408-412.
5. T, Maruyama, Y. Shiota, T. Nozaki, K. Ohta, N. Toda, M. Mizuguchi, A. A. Tulapurkar, T. Shinjo, M. Shiraishi, S. Mizukami, Y. Ando and Y. Suzuki, *Nature nanotechnol.*, 2009, **4**, 158-161.
6. M. Katayama, H. Koinuma and Y. Matsumoto, *Mater. Sci. Eng. B*, 2008, **148**, 19.
7. J. -W. Park, D. Lee, H. Kwon, S. Yoo and H. Huh, *IEEE Electron Device Lett.*, 2009, **30**, 739.
8. W. S. Shih, S. J. Young, L. W. Ji, W. Water and H. W. Shiu, *J. Electrochem. Soc.*, 2011, **158**, H609.
9. N. Zhong, H. Shima and H. Akinaga, *AIP advances*, 2011, **1**, 032167.
10. K. -H. Choi, K. -B. Chung and H. -K, Kim, *Appl. Phys. Lett.*, 2013, **102**, 153511.
11. T. Yajima, G. Oike, T. Nishimura and A. Toriumi, *Phys. Status Solid A*, 2016, **213**, 2196-2202.
12. T. Sasaki, M. Watanabe, H. Hashizume, H. Yamada and H. Nakazawa, *J. Am. Chem. Soc.*, 1996, **118**, 8329.
13. M. Osada and T. Sasaki, *Adv. Mater.*, 2012, **24**, 210.
14. L. Wang and T. Sasaki, *Chem. Rev.*, 2014, **114**, 9455-9486.
15. N. Sakai, Y. Ebina, K. Takada and T. Sasaki, *J. Am. Chem. Soc.*, 2004, **126**, 5851.
16. M. Osada, Y. Ebina, H. Funakubo, S. Yokoyama, T. Kiguchi, K. Takada and T. Sasaki, *Adv. Mater.*, 2006, **18**, 1023.
17. Y. Wang, C. Sun, X. Yan, F. Xiu, L. Wang, S. C. Smith, K. L. Wang, G. Q. Lu and J. Zou, *J. Am. Chem. Soc.*, 2011, **133**, 695.
18. M. Ohwada, K. Kimoto, T. Mizoguchi, Y. Ebina and T. Sasaki, *Sci. Rep.*, 2013, 3, 2801.
19. K. Fukuda, Y. Ebina, T. Shibata, T. Aizawa, I. Nakai and T. Sasaki, *J. Am. Chem. Soc.*, 2007, **129**, 202-209.
20. T. Sasaki, Y. Ebina, K. Fukuda, T. Tanaka, M. Harada and M. Watanabe, *Chem. Mater.*, 2002, **14**, 3524.
21. K. Nagashio, T. Nishimura, K. Kita and A. Toriumi, *Appl. Phys. Express*, 2009, **2**, 025003.
22. H. Tang, K. Prasad, R. Sanjines, P. E. Schmid and F. Levy, *J. Appl. Phys.* 1994, **75**, 2042.
23. T. Ohsaka, F. Izumi and Y. Fujiki, *J. Raman Spectroscopy*, 1978, **7**, 321.
24. O. Frank, M. Zukalova, B. Laskova, J. Kurti, J. Koltai and L. Kavan, *Phys. Chem. Chem. Phys.*, 2012, **14**, 14567.
25. T. Sasaki, *Supramolecular Sci.*, 1998, **5**, 367.
26. H. Tang, H. Berger, P. E. Schmid and F. Levy, *Solid State Communications*, 1993, **87**, 847.
27. J. Park, K. -C. Ok, B. Du. Ahn, J. H. Lee, J. -W. Park, K. -B. Chung and J. -S. Park, *Appl. Phys. Lett.*, 2011, **99**, 142104.
28. J. -M. Lee, I. -T. Cho, J. -H. Lee and H. -I, Kwon, *Appl. Phys. Lett.*, 2008, **93**, 093504.
29. D. B. Strukov, G. S. Snider, D. R. Stewart and R. S. Williams, *Nature*, 2008, **453**, 80.
30. H. Miyaoka, G. Mizutani, H. Sano, M. Omote, K. Nakatsuji and F. Komori, *Solid State Communications*, 2002, **123**, 399.
31. P. H. Wobkenberg, T. Ishiwara, J. Nelson, D. D. Bradley, S. A. Haque and T. D. Anthopoulos, *Appl. Phys. Lett.*, 2010, **96**, 082116.
32. Y. Furubayashi, N. Yamada, Y. Hirose, Y. Yamamoto, M. Otani, T. Hitosugi, T. Shimada and T. Hasegawa, *J. Appl. Phys.*, 2007, **101**, 093705.
33. R. Schroeder, L. A. Majewski and M. Grell, *Appl. Phys. Lett.*, 2004, **84**, 1004.
34. Y. Shimura, K. Nomura, H. Yanagi, T. Kamiya, M. Hirano and H. Hosono, *Thin solid films*, 2008, **516**, 5899.
35. S. Heinze, J. Tersoff, R. Martel, V. Derycke, J. Appenzeller and Ph. Avouris, *Phys. Rev. Lett.*, 2002, **89**, 106801.
36. S. Das, H. -Y. Chen, A. V. Penumatcha and J. Appenzeller, *Nano Lett.*, 2013, **13**, 100-105.
37. S. M. Sze and K. K. Ng, *Physics of semiconductor devices*, 3$^{rd}$ ed., John Wiley & Sons, Hoboken, NJ, 2007.
38. S. -L. Li, K. Komatsu, S. Nakaharai, Y. F. Lin, M. Yamamoto, X. Duan and K. Tsukagoshi, *ACS Nano*, 2014, **8**, 12836-12842.
39. K. K. Ng and R. Liu, *IEEE Trans. Electron Devices*, 1990, **37**, 1535-1537.
40. N. Zhong, H. Shima and H. Akinaga, *Appl. Phys. Lett.*, 2010, **96**, 042107.

# Field effect transistor of thin anatase fabricated through solid state transformation of titania nanosheet

Shunya Sekizaki[a], Minoru Osada[b] and Kosuke Nagashio[a,c]*

[a]Department of Materials Engineering, The University of Tokyo, Tokyo 113-8656, Japan
[b]National Institute of Materials Science, Ibaraki 305-0044, Japan
[c]PRESTO, Japan Science and Technology Agency (JST), Tokyo 113-8656, Japan
*E-mail: nagashio@material.t.u-tokyo.ac.jp

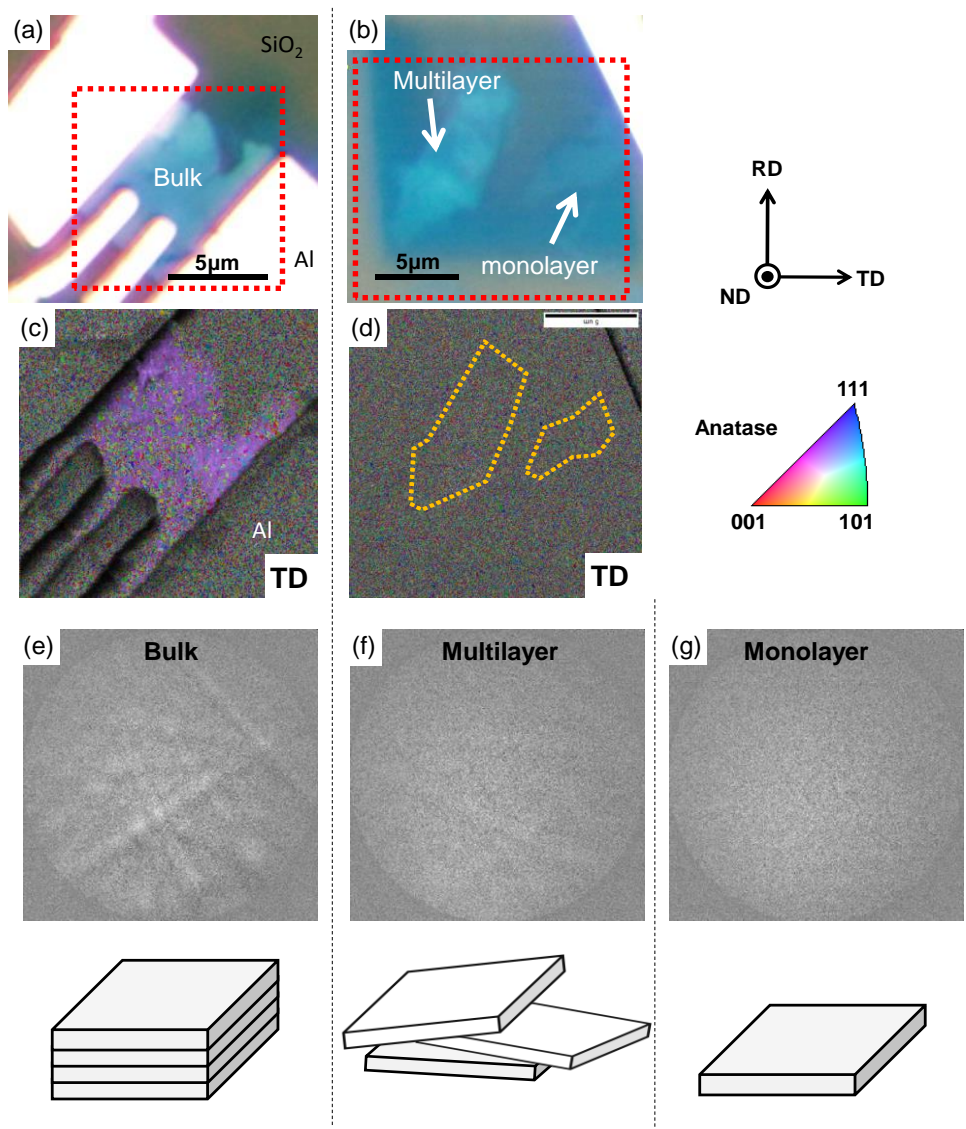

**Fig. S1**. EBSD analysis for three kinds of anatase samples. (a,b) Optical micrograph of three kinds of anatase samples. (c,d) EBSD orientation maps of the transverse direction (TD) are colored using the inverse pole figure triangle. (e-g) Kikuchi patterns observed

at an arbitrary point of samples. The same Kikuchi pattern can be seen for different positions for bulk anatase sample, while the different Kikuchi patterns are observed for multilayer anatase sample. For monolayer, Kikuchi pattern is not evident.

Based on this EBSD analysis, "Bulk" is single crystalline nature, because each $Ti_{0.87}O_2$ nanosheet is stacked with retaining the crystal orientation (actually, $Ti_{0.87}O_2$ nanosheet was not exfoliated in the $Ti_{0.87}O_2$/TBA suspension.) "Multilayer" is the sample in which $Ti_{0.87}O_2$ nanosheets are randomly laminated without crystal orientation matching. Therefore, "multilayer" shows the polycrystalline nature from EBSP data. In the present definition for "multilayer" and "bulk", the layer number is not taken into consideration. However, the thickness for bulk (~10-15 nm) is generally thicker than that for multilayer (~3-10 nm).